\documentclass[aps,prr,twocolumn,groupedaddress]{revtex4-1}

\usepackage{graphicx,bbm,bm,hyperref}  
\usepackage{amsmath}
\usepackage{amsthm}

\usepackage[pdftex]{color}

\def\sx#1{\sigma^{\rm x}_{#1}}
\def\sy#1{\sigma^{\rm y}_{#1}}
\def\sz#1{\sigma^{\rm z}_{#1}}

\def\Z2{\mathbbm{Z}_2}

\def\tr#1{{\rm tr}(#1)}
\def\1{\mathbbm{1}}
\def\ket#1{{| #1 \rangle}}
\def\bra#1{{\langle #1 |}}
\def\braket#1#2{{\langle #1 | #2 \rangle}}

\def\tit#1{{\em #1},}

\begin{document}

\title{Solvable non-Hermitian skin effect in many-body unitary dynamics}

\author{Marko \v Znidari\v c}
\affiliation{Physics Department, Faculty of Mathematics and Physics, University of Ljubljana, 1000 Ljubljana, Slovenia}

\date{\today}

\begin{abstract}
We study unitary evolution of bipartite entanglement in a circuit with nearest-neighbor random gates. Deriving a compact non-unitary description of purity dynamics on qudits we find a sudden transition in the purity relaxation rate the origin of which is in the underlying boundary localized eigenmodes -- the skin effect. We provide the full solution of the problem, being one of the simplest iterations of two-site matrices, namely, that each is a sum of only two projectors. This leads to rich dynamics influenced by the Jordan normal form of the kernel and, most importantly, a spectrum that is completely discontinuous in the thermodynamic limit. It provides a simple example of how a seemingly innocuous many-body unitary evolution can harbor interesting mathematical effects: an effective non-symmetric Toeplitz transfer matrix description causes a phantom relaxation, such that the correct relaxation rate is not given by the matrix spectrum, but rather by its pseudospectrum.
\end{abstract}

\maketitle

\section{Introduction}
Success of physics relies on managing to describe seemingly complicated phenomena in a simple way. Trying to understand behavior of systems of many particles has been an important theme ever since the exact solution of the two-body problem of celestial mechanics. As the number of particles increases the resulting many-body physics is in general not analytically solvable anymore. Unitary evolution in such large Hilbert space in quantum mechanics, or unitary Koopman propagation in classical phase space~\cite{Braun}, is simply too complicated. Nevertheless, in the thermodynamic limit (TDL) when the number of particles goes to infinity things can simplify for sufficiently well-behaved quantities -- the driving principle of Statistical physics.

The TDL of system size $n \to \infty$ together with the long time limit $t \to \infty$ can bring in the game some very interesting effects that are at first sight seemingly at odds with unitarity. For instance, the overlap preservation due to unitarity would seem to prohibit exponential sensitivity to initial conditions -- a characteristic of chaos, or, reversibility of unitary evolution is against the entropy increase of the 2nd law of thermodynamics. A resolution of such fallacies often relies on (i) the fact that we are not interested in unitary evolution on the full space, particularly in many-body systems, but rather in the evolution of few relevant quantities obtained by either coarse graining or by integrating out most degrees of freedom, and (ii) the fact that limits $n \to \infty$ and $t \to \infty$ often do not commute. Behavior at fixed $t$ and $n \to \infty$ (the correct TDL) can be counterintuitive and against our expectation based on unitary dynamics in the limit $t \to \infty$ at fixed $n$, a fact that will feature also in our study.

Deriving and understanding such effective description of relevant observables is therefore important. In the present work we are going to derive an exact description for a specific quantity in a many-body quantum system, resulting in a non-Hermitian matrix that will cause rather interesting effects, for instance, the rate of generating entanglement will exhibit a sudden change with time. Exact solution will bring under the same roof mathematics related to a number of recent interesting observations like topology-induced non-Hermitian skin effect~\cite{Lee16,Torres18,Wang,Shen18,Kunst18,Lee19,Kawabata19,Gong20,Slager20,Okuma20,Zhang20} (for reviews see Refs.~\cite{Torres20,Kunst21}), disparity between Lindbladian relaxation time~\cite{Mori20,Mori21,Ueda21} and the inverse gap~\cite{PRE15}, metastable Majorana bosons localized at a boundary of a Lindbladian system~\cite{Viola21}, and a phantom two-step relaxation~\cite{PRX21,PRR22}. A solvable instance of the many-body non-Hermitian skin effect that exactly emerges from the underlying unitary dynamic, rather than starting with a non-Hermitian Hamiltonian, should be of value. We also provide a solvable example of the recently numerically observed phantom relaxation of purity~\cite{PRX21}, seen also in out-of-time-ordered correlation (OTOC) functions~\cite{PRR22} and von Neumann entropy, and occurring in a variety of different random circuits. While exact solutions are always special, and in this sense not generic, we should stress that our results present a solvable example of this more generic phenomenon, thereby providing an explicit understanding of the underlying inner workings. It also adds a new analytic result about entanglement in random circuits; see Refs.~\cite{Oliveira,PRA08,Harrow09,Viola10,Zanardi12,Adam18,Frank18,Chan18,Zhou19,Kuo20,Bruno20} for a set of previous exact results.

All this physics is a manifestation of rich underlying mathematics. While non-Hermitian systems with their exceptional points have been studied extensively, see e.g. reviews~\cite{Bender,Ueda,Heiss}, focus has been on the single-particle setting (see though e.g. Ref.~\cite{Luitz19} for a many-body case). The many-body setting brings with it an additional TDL. As we will see, the non-Hermitian Toeplitz matrices~\cite{Bottcher} are such that their spectrum for any finite size is completely different than the spectrum of the operator in the TDL. While such mathematical properties have been observed and appreciated in numerical analysis~\cite{Trefethen}, e.g. in the convergence rate of algorithms like Gauss-Seidel iteration, they are less known in physics, see though Ref.~\cite{Viola21} for some cases. Also, the elementary step of our effective description is rather simple: it is just a sum of two projectors whose action is a sequence of project-rotate-stretch steps. Understanding elementary building blocks has proved beneficial in the past, an example is the stretch-and-fold scenario of classical chaos as epitomized by the horseshoe map~\cite{Ott}.

\section{Average purity evolution via transfer matrix}

We want to study bipartite entanglement in a chain of $n$ qudits with local Hilbert space dimension $d$. Unitary propagator $U$ for one unit of time is a product of independent identically distributed (i.i.d.) 2-site random gates $U_{j,j+1}$ distributed according to the unique unitary-invariant Haar measure (see e.g. Ref.~\cite{Karol}),
\begin{equation}
  U=U_{n-1,n} U_{n-2,n-1} \cdots U_{1,2}.
  \label{eq:U}
\end{equation}
Gates are therefore applied in a staircase configuration with open boundaries, see e.g. Refs.~\cite{Adam18,PRX21,Green21} for cases where such configuration has been discussed. The pure state at time $t$ is $\ket{\psi(t)}=U^t \ket{\psi(0)}$. For exact analytical derivations we will require that all 2-site gates are i.i.d. random gates (in total $(n-1)t$ i.i.d. gates in the whole $U^t$), however, numerically we will see that in the TDL physics is the same even if we apply the same gate at all sites and all times (a single Haar random gate in the whole $U^t$). Purity for a bipartition into 1st $k$ sites (subsystem A) and the rest (subsystem B) is
\begin{equation}
  I_k(t)=\tr{\rho_k^2(t)},\quad \rho_k(t)={\rm tr}_{k+1,\ldots,n}(\ket{\psi(t)}\bra{\psi(t)}).
  \label{eq:Ik}
\end{equation}
For a separable initial state one has $I_k(0)=1$ after which purity decays, i.e., entanglement increases. At long time it will converge to $I_k(\infty)=(d^k+d^{n-k})/(1+d^n)$, being purity of a random state~\cite{Lubkin}. We are interested in the decay of purity to this asymptotic value.

One can show that the average dynamics of the squares of expansion coefficients of $\rho(t)$ is governed by a Markovian process~\cite{Oliveira,DahlstenJPA} acting on the operator space of dimension $d^{2n}$. The dimensionality can though be reduced down to dimension $2^n$~\cite{PRA08}, regardless of $d$, with the Markovian matrix being equal to the Hamiltonian of a XY spin chain. More recently an equivalent mapping has been obtained by an independent and more general means~\cite{Kuo20} and we shall use that formulation (see Appendix~\ref{App:A} for technical details and different possible representations). Defining purities for all $2^n$ possible bipartitions by $I_\mathbf{s}(t)$, where the $n$-bit string $\mathbf{s}=(s_1,s_2,\ldots,s_n)$ encodes the bipartition $A+B$, $s_j=1$ denoting the $j$-th site being in A, while $s_j=0$ denotes it is in B (we will use a scalar index $k$ in $I_k$ to denote a bipartition with the first $k$ contiguous sites in A, and a vector index $\mathbf{s}$ like in $I_\mathbf{s}$ for an arbitrary bipartition), one can show~\cite{Kuo20} that under a random 2-site unitary the average $I_\mathbf{s}(t+1)$ is obtained from $I_\mathbf{s}(t)$ simply by a matrix multiplication,
\begin{equation}
I_\mathbf{s}(t+1)= \sum_{\mathbf{s}'} [M]_{\mathbf{s},\mathbf{s}'} I_{\mathbf{s}'}(t),
\label{eq:iter}
\end{equation}
where a $2^n\times 2^n$ matrix $M$ is a product of 2-site matrices $M_{j,k}$ describing individual gates applied in a given protocol per unit of time, in our case for the staircase configuration
\begin{equation}
  M=M_{n,n-1}\cdots M_{1,2},\qquad M_{j,k}=\begin{pmatrix}
		1 & 0 & 0 & 0 \\
		\alpha & 0 & 0 & \alpha \\
		\alpha & 0 & 0 & \alpha \\
		0 & 0& 0& 1
  \end{pmatrix},
  \label{eq:M}
\end{equation}
where the 2-site $M_{j,k}$ acts nontrivially only on sites $j$ and $k$ and is written in the basis $s_js_k$ ordered as $\{00,10,01,11 \}$, while for $\alpha$ one gets~\cite{Kuo20} 
\begin{equation}
\alpha=\frac{d}{d^2+1}.
\end{equation}
We therefore have (\ref{eq:iter}) a transfer matrix description of the time evolution of purity averaged over random unitaries. Starting with a product initial state, for which all purities are $I_\mathbf{s}(0)=1$, i.e., the initial purity vector is $(1,\ldots,1)$, we obtain the average purity at time $t$ by a multiplication by $M^t$.

\begin{figure}[t!]
\centerline{\includegraphics[width=.9\columnwidth]{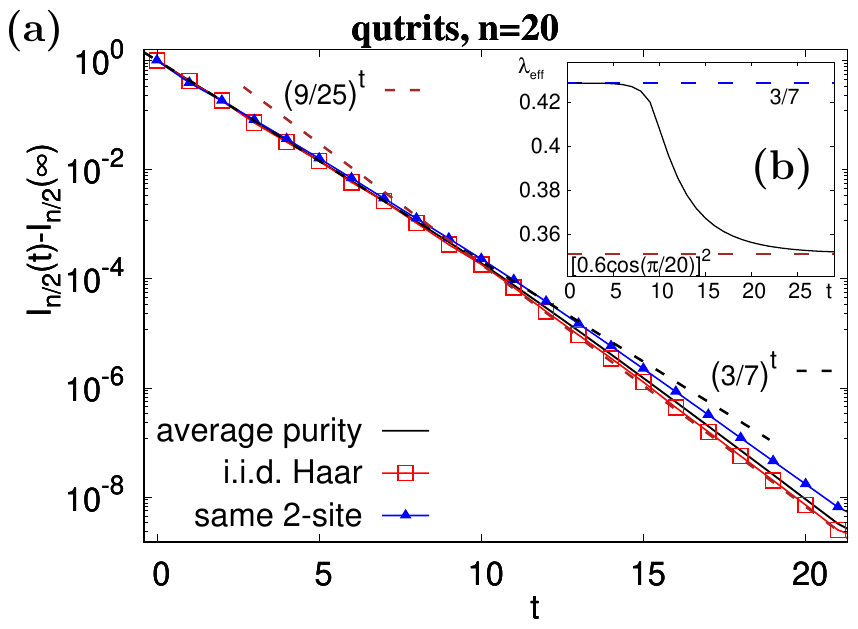}}
\caption{(Color online) Purity decay for a single random circuit realization with $20$ qutrits and a half-half bipartition (squares and triangles). Due to a self-averaging in large Hilbert space ($d^n \approx 3\cdot 10^9$) results are essentially the same as for the exact average purity (full curve). Brown dashed curve is the theoretical asymptotic decay given by the transfer matrix eigenvalue, black dashed line is theoretical phantom decay. The inset shows the transition in the local $\lambda_{\rm eff}$ defined as $I(t) = C \lambda_{\rm eff}^t$.}
\label{fig:u9n20}
\end{figure}
Before going to exact derivations let us have a peek at how purity looks like so that it will be clear what interesting physics we are trying to explain. We show in Fig.~\ref{fig:u9n20} with squares an example of purity evolution for a single qutrit ($d=3$) circuit realization with all 2-site gates being independent random (as well as for the case when all 2-site gates are equal to the same random unitary, blue triangles). The main observation is that initially ($t<10$ for the shown $n=20$) the decay is exponential $I(t)=(3/7)^t$, whereas at later time it transitions into $I(t) \sim (9/25)^t$ (see also the inset that shows transition in $\lambda_{\rm eff}$ defined by $\lambda_{\rm eff}(t)=\exp{[I'(t)/I(t)]}$). What is surprising is that $\lambda_{\rm ph}=3/7$ is larger than any nontrivial eigenvalue of $M$. Namely, $M$ has an eigenvalue $1$ corresponding to the steady state with purity $I(\infty)$, while the 2nd largest eigenvalue is $\lambda_2=9/25$. One would expect the asymptotic decay $I(t) \sim  \lambda_2^t$, while in fact in the TDL the transition time between the two decays moves to infinity and one observes decay with $\lambda_{\rm ph}=3/7$. Such sudden transition in the relaxation rate has been recently numerically observed in purity~\cite{PRX21} and OTOCs decay~\cite{PRR22} in a number of different qubit circuits. We will explain it by providing an exact solution for any $d$ for our specific circuit.

\begin{figure}[t!]
\centerline{\includegraphics[width=.7\columnwidth]{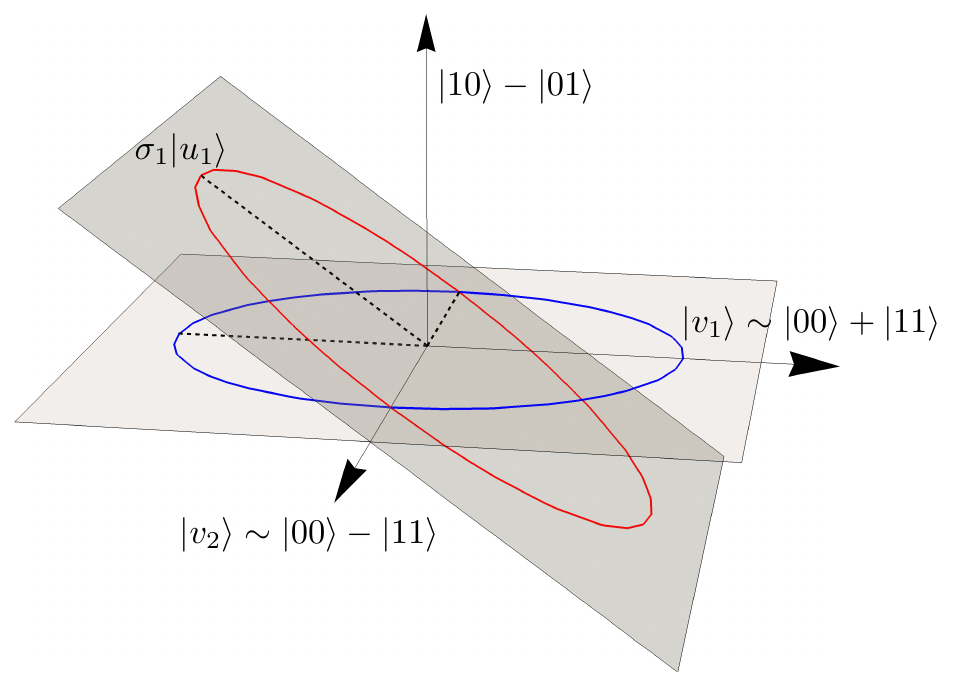}}
\caption{(Color online) Action of the 2-site $M_{j,k}$ (\ref{eq:M}): project to a 2-dimensional plane $\ket{v_{1,2}}$, then rotate into plane $\ket{u_{1,2}}$, and stretch along $\ket{u_1}$ (the blue circle is transformed to a red ellipse).}
\label{fig:rotacija}
\end{figure}

\section{Results}

\subsection{Two-site transfer matrix}
Let us begin by making a few observations on $M$ whose properties we need to understand. It is a product of quite simple 2-site matrices $M_{j,k}$ (\ref{eq:M}) that have only two nonzero eigenvalues equal to $1$ and therefore only two projectors in the spectral decomposition, see App.~\ref{App:A}. 

Singular value decomposition (SVD) also has just two nonzero singular values $\sigma_1=\sqrt{1+4\alpha^2}$ and $\sigma_2=1$, $M_{j,k}=\sigma_1 \ket{u_1}\bra{v_1}+\ket{u_2}\bra{v_2}$, where (upto normalization) $u_1=(1,2\alpha,2\alpha,1)$, $u_2=(1,0,0,-1)$, and $v_1=(1,0,0,1)$, $v_2=u_2$. Vectors $u_j$ and $v_j$ are orthonormal, $\braket{u_j}{u_k}=\delta_{j,k}$, $\braket{v_j}{v_k}=\delta_{j,k}$, offering a nice geometrical interpretation (Fig.~\ref{fig:rotacija}): from a 4-dimensional 2-site space we (i) project to a 2-dimensional subspace spanned by basis $\{\ket{v_1},\ket{v_2}\}$, then (ii) rotate to a plane spanned by $\{\ket{u_1},\ket{u_2}\}$, and finally (iii) stretch by a factor $\sigma_1>1$ along $\ket{u_1}$. The whole circuit $M$ is an iteration of such elementary project-rotate-stretch steps on successive sites. Such an elementary step could perhaps serve as a simple nontrivial toy unit-step to describe certain features of non-unitary dynamics. 

\subsection{Exact purity} 
Spectral properties of the full many-body $M$ though do not follow in any simple way from those of $M_{j,k}$. While one can infer some of its properties (see App.~\ref{App:A}) we will take a different route and derive a direct evolution equation for $n-2$ purities $I_k(t)$, $k=2,\ldots,n-1$ (\ref{eq:Ik}). Remember, those are purities for a subsystem A being the first $k$ sites. The simplifying property is that for e.g. $k=n/2$ we need $I_{1\cdots1 0\cdots 0}(t)$, i.e., all last $n/2$ bits in $\mathbf{s}$ have to be 0, and one can reach the $00$ state by $M_{j,k}$ only from the $00$ (1st row of $M_{j,k}$ in Eq.(\ref{eq:M})). For the staircase configuration one can then work backwards in time and see from which terms at previous times one can get a given bit string. It turns out, see App.~\ref{App:B} , that the equations are
\begin{equation}
  I_k(t)=\alpha^k+\sum_{r=2}^{k+1} \alpha^{k+2-r} I_r(t-1).
  \label{eq:I}
\end{equation}
This recursion can be used to get exact expressions for purities, starting with $I_k(0)=1$. With increasing time expressions get more and more complicated (see App.~\ref{App:B}), however, all are of the form
\begin{equation}
  I_k(t)=\left( \frac{\alpha}{1-\alpha}\right)^t +\alpha^k p_{t-1}(n),
\end{equation}
where $p_r(x)$ is a polynomial of degree $r$ in $x$ (polynomial also depends on $k$). Fixing extensive subsystem A with $k \propto n$, for instance $k=n/2$ for the half-half bipartition, letting $n \to \infty$ the 2nd term will be $\alpha^k \to 0$ ($\alpha<1/2$ for $d>1$), leaving just
\begin{equation}
I_{k}(t)=\lambda_{\rm ph}^t, \qquad \lambda_{\rm ph}=\frac{\alpha}{1-\alpha}=\frac{d}{d(d-1)+1}.
\label{eq:Iph}
\end{equation}
For $d=3$ this gives $\lambda_{\rm ph}=3/7$ seen in Fig.~\ref{fig:u9n20}, and explains the phantom eigenvalue $\lambda_{\rm ph}=2/3$ numerically observed for $d=2$ in Ref.~\cite{PRX21}.

An alternative and more insightful route is via spectral properties of $M$. The recursion (\ref{eq:I}) can be compactly written as a matrix iteration of the purity vector $\mathbf{I}=(I_2,I_3,\ldots,I_{n-1})$,
\begin{equation}
\mathbf{I}(t+1)=\mathbf{a}+T \mathbf{I}(t), 
\end{equation}
where $\mathbf{a}=(\alpha^2,\alpha^3,\ldots,\alpha^{n-2},\alpha^{n-1}+\alpha)$, and the Toeplitz matrix $T$ (a matrix that has the same matrix elements along each diagonal):
\begin{equation}
T=\begin{pmatrix}
    \alpha^2 & \alpha & 0 & \cdots & 0 \\
    \alpha^3 & \alpha^2 & \alpha & \cdots & 0\\
    \vdots   &  \vdots & \ddots & \cdots & \vdots \\
    \alpha^{n-2} & \alpha^{n-3} & \ddots & \ddots & \alpha\\
    \alpha^{n-1} & \alpha^{n-2} & \cdots & \cdots & \alpha^2 
\end{pmatrix}.
\label{eq:T}
\end{equation}
With $T$ we are describing a subset of all $2^n$ bipartitions, and therefore eigenvalues of $T$ are also eigenvalues of $M$. In fact, the largest eigenvalues of $M$ are precisely those of $T$. The matrix $T$ has a lower-Hessenberg form (i.e., nonzero lower triangle plus a nonzero superdiagonal) and its spectral decomposition can be written out explicitly in terms of Chebyshev polynomials (App.~\ref{App:C} and Ref.~\cite{Fairweather71}). It has a degenerate kernel of dimension $\frac{n}{2}-1$ with a single Jordan normal block~\cite{Weintraub}, and $\frac{n}{2}-1$ nondegenerate eigenvalues $\tilde{\lambda}_j$,
\begin{equation}
\tilde{\lambda}_j =4\alpha^2 \cos^2{\varphi_j}  ,\quad \varphi_j=\frac{j\pi}{n},\quad j=1,\ldots,\frac{n}{2}-1.
\label{eq:lj}
\end{equation}
The largest eigenvalue of $T$, which is also the 2nd largest eigenvalue $\lambda_2$ of $M$, is therefore in the TDL
\begin{equation}
  \lambda_2=4\alpha^2=\frac{4d^2}{(d^2+1)^2}.
\end{equation}
For $d=3$ it is $\lambda_2=9/25$ (giving the asymptotic decay in Fig.~\ref{fig:u9n20}). We have therefore analytically shown that at fixed $t$ and $n \to \infty$ the decay is $I_{n/2}(t)-I_{n/2}(\infty)=\lambda_{\rm ph}^t$, while at fixed $n$ and sufficiently long time one will have $I_{n/2}(t)-I_{n/2}(\infty) \sim \lambda_{2}^t$. One can also see (App.~\ref{App:C}) that for extensive $k$ the asymptotic decay $\lambda_2^t$ in $I_k(t)-I_k(\infty)$ will kick-in only at a time that is proportional to $n$. Therefore, at any fixed $t$ and for $n\to \infty$ the decay will be (\ref{eq:Iph}) as if there would be a phantom eigenvalue $\lambda_{\rm ph}$ in the spectrum of $M$ ($\lambda_1=1 > \lambda_{\rm ph} > \lambda_2$). This is the so-called phantom relaxation~\cite{PRX21}. How can that be? 

\subsection{Spectral resolution}
On the level of the spectral decomposition, where we would write a matrix in terms of its eigenvalues and left and right eigenvectors, it happens because the left and right eigenvectors of $T$ are exponentially localized. This leads to expansion coefficients of the initial vector over left eigenvectors of $T$ that grow exponentially with $n$, thereby delaying the appearance of $\asymp \lambda_2^t$ to later and later times as $n$ increases. 

\begin{figure}[t!]
\centerline{\includegraphics[width=.95\columnwidth]{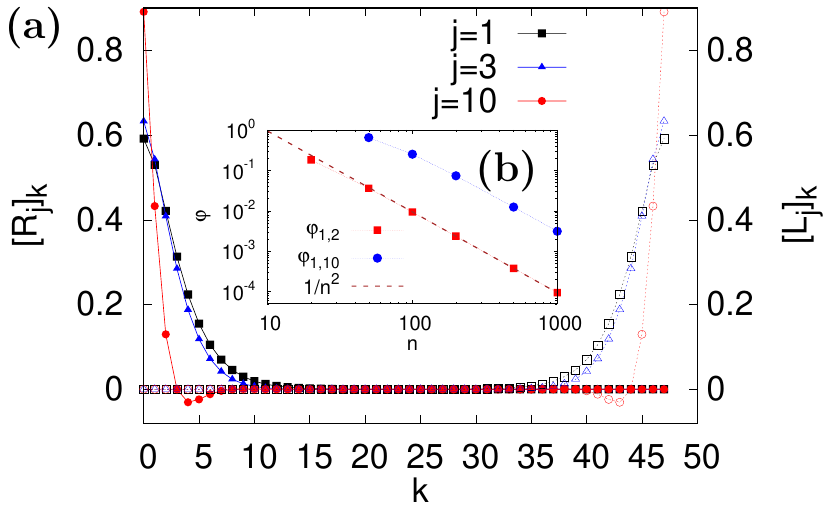}}
\caption{(Color online) The skin effect of eigenvectors of $T$ (\ref{eq:T}) for $d=3$. (a) Shown are 3 right (full symbols) and left eigenvectors (empty symbols), all showing localization at the left or right edge, respectively. (b) The angle $\varphi_{1,j}$ between eigenvectors $R_1$ and $R_j$ falls off as $\sim 1/n^2$.}
\label{fig:lastnid3}
\end{figure}

Denoting by $[R_j]_k$ the unnormalized $k$-th component ($k=1,\ldots,n-2$) of the right eigenvector of $T$ corresponding to $\tilde{\lambda}_j$, one has
\begin{equation}
[R_j]_k= (2\alpha \cos{\varphi_j})^{k-2} \frac{\sin{[(k+1)\varphi_j]}}{\sin{\varphi_j}},
\end{equation}
while the left eigenvectors are simply the reflected $R_j$,
\begin{equation}
[L_j]_k= (2\alpha \cos{\varphi_j})^{n-3-k} \frac{\sin{[(n-k)\varphi_j]}}{\sin{\varphi_j}}.
\end{equation}
Few eigenvectors are shown in Fig.~\ref{fig:lastnid3}. One can see that all eigenvectors are localized at the boundary, resulting in overlaps of unnormalized $\braket{L_j}{R_j}$ being exponentially small in $n$. What is more, an increasing number of left (as well as right) eigenvectors is becoming almost co-linear (the inset). With increasing $n$ those eigenvectors and eigenvalues are becoming essentially degenerate. These properties are responsible for $\lambda_2$ not giving the correct decay in the TDL. Our solution provides an example of a non-Hermitian skin effect -- a situation where an extensive number of eigenmodes becomes localized at a boundary~\cite{Torres20,Kunst21}, and which is related to topological invariants (similar to a winding number of Toeplitz matrices~\cite{Bottcher}). The same effect is here responsible for the transition in the purity relaxation rate. Using the spectral decomposition of $T$ one can explicitly write down $I_k(t)$ in terms of a sum, from which another interesting feature arises: using only nonzero $\tilde{\lambda}_j$ one will not correctly describe $I_{n/2}(t)$ for $t<n/4$; to get that short-time decay correctly one needs to properly account also for the Jordan structure of the kernel, see App.~\ref{App:C}. To discuss the phantom relaxation itself though, the Jordan structure of the kernel is not crucial.

\subsection{Noncommutativity of limits}
We see that in such phantom relaxation the limits $t \to \infty$ and $n \to \infty$ do not commute. On the other hand the spectrum of finite $T$ (or $M$) is perfectly smooth -- nothing special happens with $\tilde{\lambda}_j$ as one increases $n$, see Eq.(\ref{eq:lj}). Resolution of this ``paradox'' lies in the difference between the spectrum of a finite Toeplitz matrix, and the spectrum of the corresponding Toeplitz operator (an infinite matrix $T_\infty$). 

\begin{figure*}[t!]
\centerline{\includegraphics[width=.62\columnwidth]{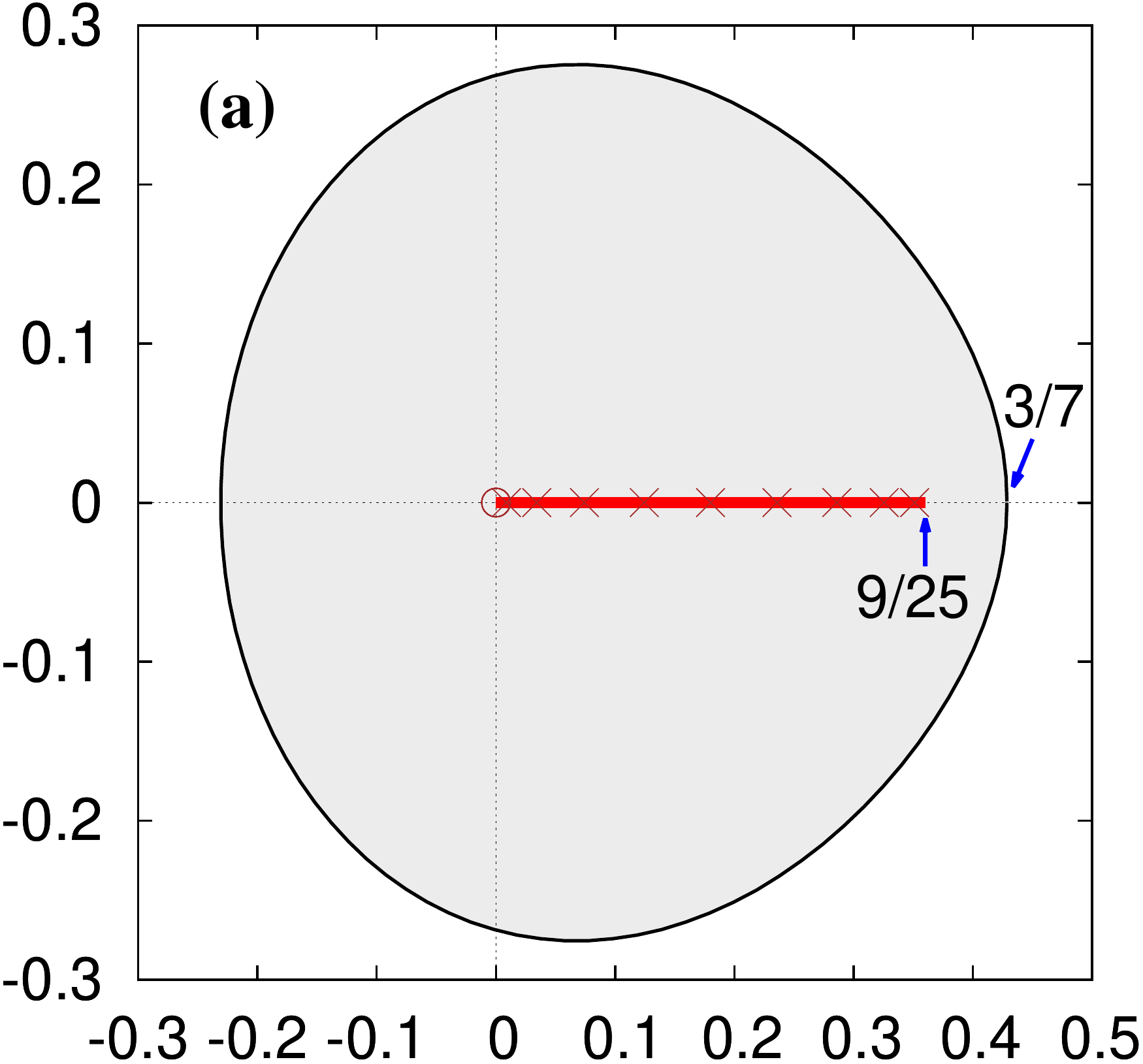}\hskip3mm\includegraphics[width=.62\columnwidth]{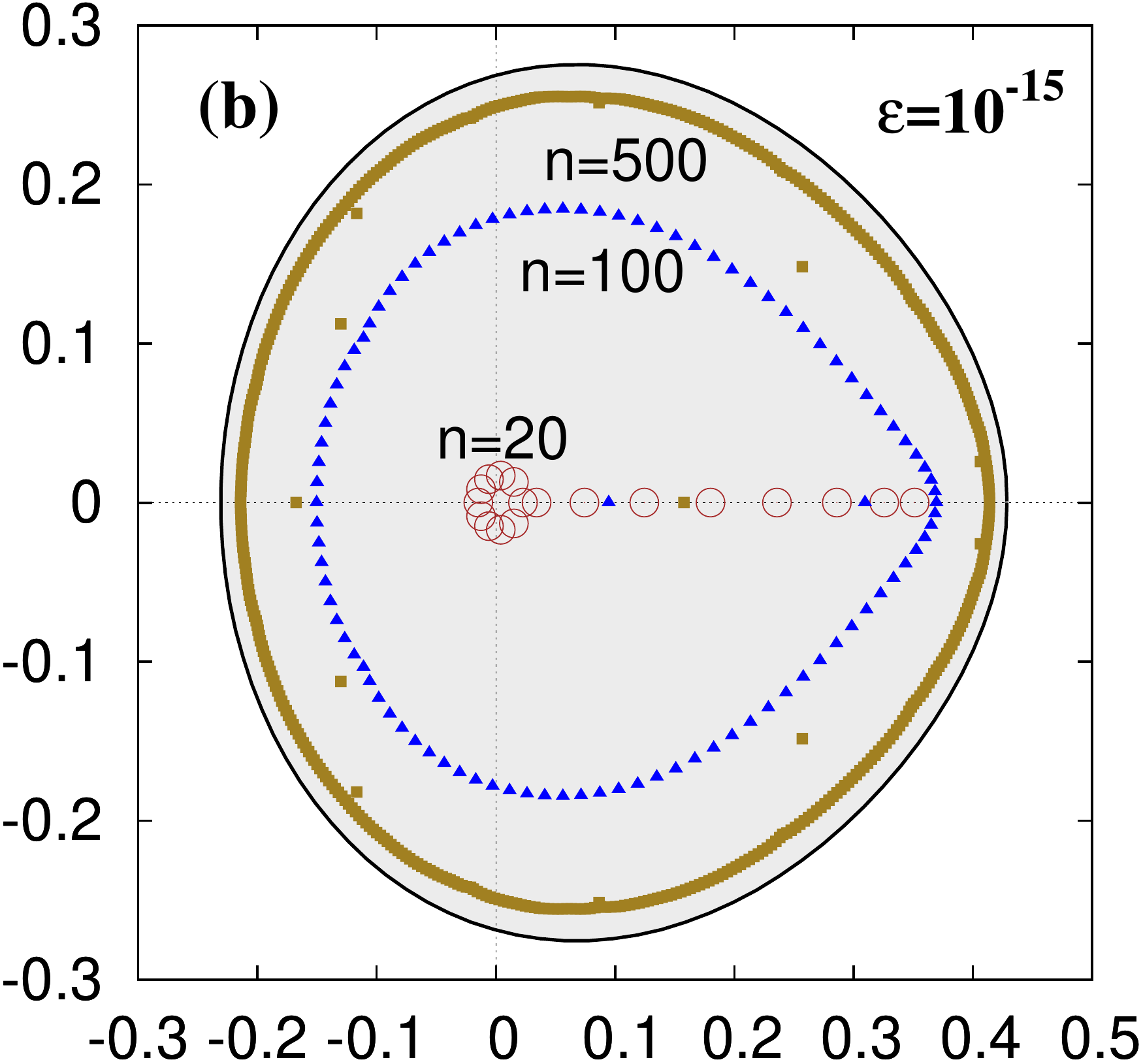}\hskip3mm\includegraphics[width=.62\columnwidth]{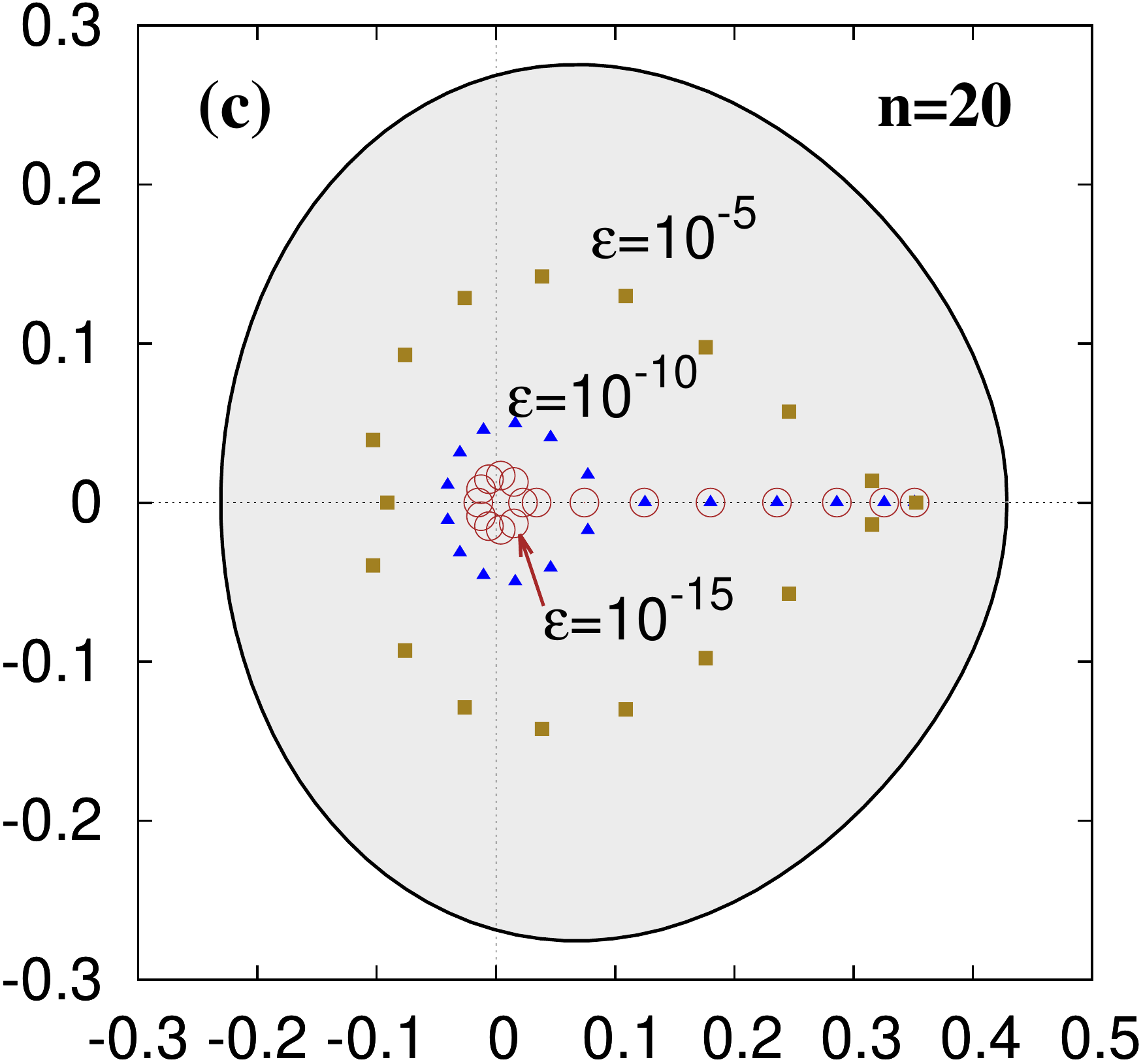}}
\caption{(Color online) (a) Exact spectrum of the Toeplitz operator $T_\infty$ (shaded oval) vs. the spectrum of a finite large Toeplitz matrix $T$ (red line) for $d=3$. Red symbols show eigenvalues for small $n=20$ (crosses $\tilde{\lambda}_j$, circle $\lambda=0$). (b) Pseudospectrum: numerically computed spectrum of a single $T+\varepsilon E$ ($E$ is a matrix of i.i.d. real Gaussian random numbers; $\varepsilon=10^{-15}$) for $n=20, 100, 500$ (circles, triangles, and squares, respectively). (c) Decreasing perturbation $\varepsilon$ at fixed $n=20$ the spectrum of $T+\varepsilon E$ (points) goes towards that of a finite matrix $T$ (\ref{eq:lj}).}
\label{fig:spekter}
\end{figure*}
Toeplitz matrices~\cite{Bottcher} can be compactly specified in terms of the so-called symbol $a(z=e^{i \theta})$, being the Fourier transformation of its diagonals. That is, we can write $T_{i,j}=a_{i-j}$, where $a_k=\frac{1}{2\pi}\int_0^{2\pi}{a(e^{i\theta}) e^{-i k \theta}d\theta}$ is defined in terms of the symbol. For our specific $T$ one can easily calculate the symbol, and it is
\begin{equation}
  a(z)=\frac{\alpha}{z}+\frac{\alpha^2}{1-\alpha z}.
  \label{eq:symbol}
\end{equation}
It is a sum of $\alpha/z$ giving the superdiagonal, and $\alpha^2/(1-\alpha z)$ responsible for the lower-triangle. Many powerful results are known for Toeplitz matrices~\cite{Bottcher}, one of them stating (under certain conditions fulfilled by $T$) that the spectrum of the Toeplitz operator $T_\infty$ is given by the set $a(z)$ for $|z|\le 1$. This set is the oval shape visible in Fig.~\ref{fig:spekter}. We can see that the spectrum of $T$ is completely discontinuous: for any finite $n$ the spectrum is real, $\tilde{\lambda}_j \in [0,\lambda_2]$, and is nothing like the spectrum of $T_\infty$, which fills the lemon-shaped figure given by the symbol. In particular, the norm of $T_{\infty}$ is equal to $\Vert a \Vert_\infty=a(z=1)=\alpha/(1-\alpha)$. The phantom eigenvalue $\lambda_{\rm ph}$ that determines purity decay in the TDL, and was mysteriously absent from the spectrum of $T$, is simply the norm of $T_\infty$,
\begin{equation}
\lambda_{\rm ph}=\Vert a \Vert_\infty=\alpha/(1-\alpha).
\end{equation}
 
Such discontinuities are well known in the theory of Toeplitz matrices and are for instance important in numerical analysis when analyzing stability and convergence rate of algorithms. In fact, the matrix $T$ (\ref{eq:T}) is for $\alpha=\frac{1}{2}$ very similar to the iteration matrix of the Gauss-Seidel method for solving systems of linear equations~\cite{Trefethen92}. When dealing with such ``fragile'' spectra what usually matters is not the spectrum, but rather the pseudospectrum~\cite{Trefethen}. A pseudospectrum is defined as the spectrum of a perturbed matrix. An $\varepsilon$ pseudospectrum ${\rm sp}_\varepsilon(z)$ of $T$ is a set of $z$ such that $z$ is in the spectrum of $T+\varepsilon E$ for some $\Vert \varepsilon E \Vert< \varepsilon$. For normal matrices (e.g. Hermitian or unitary) the pseudospectrum is essentially the same as the spectrum, for non-normal ones this is not necessarily the case. A quick estimate of the pseudospectrum can be obtained by perturbing $T$ by a single matrix, and this is shown in Fig.~\ref{fig:spekter}. We can see that fixing small $\varepsilon$ and letting $n \to \infty$ the pseudospectrum indeed goes to that of $T_\infty$ (Fig.~\ref{fig:spekter}(b)), rather than to that of a finite matrix. Beware also that the limits $\varepsilon \to 0$ and $n \to \infty$ do not commute: if one decreases $\varepsilon$ at fixed $n$ the spectrum goes to that of a finite matrix (Fig.~\ref{fig:spekter}(c)). 

The pseudospectrum is smooth in $n$, including the TDL, while the spectrum is not. Looking at the pseudospectrum of $T$ instead of at the spectrum would therefore even for finite $n$ predict the correct relaxation rate $\lambda_{\rm ph}$. We have an example where being exact (exact finite $T$) is actually wrong -- it gives $\lambda_2$ that predicts an incorrect relaxation rate, while being slightly wrong (perturbed $T$) is correct -- pseudospectrum with $\lambda_{\rm ph}$ gives the correct relaxation rate. The importance of the pseudospectrum has recently been stressed in several situations~\cite{Okuma20b,Viola21,Okuma21,Yoshida21}.

\section{Conclusion}

Calculating purity dynamics in a unitary staircase configuration random circuit on qudits reveals a transition in the entanglement growth rate due to localized eigenmodes (skin-effect) of the underlying non-symmetric matrix describing purity dynamics. Explicit results for eigenvalues and eigenvectors allow us to get, among other things, exact results for the initial phantom relaxation rate, i.e. the one relevant in the TDL, and the asymptotic relaxation rate given by the 2nd largest eigenvalue of the transfer matrix as observed in finite systems. The transition time between the two rates diverges in the TDL, while their ratio increases for larger qudit dimensions.

All this is a consequence of rich mathematics of the underlying solvable Toeplitz matrix, which has an extensively large Jordan normal form kernel, the main point though is that its spectrum discontinuously changes in the TDL. Any finite matrix has a completely different spectrum than the infinite-size matrix. The relevant quantity for purity dynamics in the TDL is in fact not the spectrum but rather the pseudospectrum.

Similar techniques as applied here could be applied to other circuits with 2-qudit random gates (for a brick-wall example see App.~\ref{App:C}). Considering that a similar phantom relaxation has been found for many other random circuits with single or 2-site random gates, as well as for out-of-time-ordered correlations, a natural question is whether the physical/mathematical reason there is also the same, i.e., the underlying non-Hermitian Toeplitz-like matrix with a discontinuous spectrum. Classifying behavior of products of other simple 2-site transfer matrices that are sums of projectors would also be desirable. For sums instead of products of elementary projectors a full classification of ground state physics has been achieved~\cite{Bravyi15}. That would help in understanding in which other non-unitary situations the phenomenon occurs. Having a solvable example of a many-body non-Hermitian skin effect that naturally emerges from an exact underlying unitary dynamics, rather than starting with a non-Hermitian Hamiltonian put in by hand, should in itself be of interest.

I would like to acknowledge support by Grants No.~J1-1698 and No.~P1-0402 from the Slovenian Research Agency.

\appendix

\section{Different formulations of purity evolution}
\label{App:A}

\subsection{Propagating state}

Writing $\rho(t)$ in terms of a local basis one can show that the evolution of average squares of expansion coefficients is given by a Markovian process~\cite{Oliveira,DahlstenJPA}: a random gate $U_{j,j+1}$ maps $\mathbf{1}\otimes \mathbf{1}$ to $\mathbf{1}\otimes \mathbf{1}$ (trace preservation), while any other product $\sigma^\alpha \otimes \sigma^\beta$ is mapped equiprobably to any other of 15 of the same products (using qubits $d=2$ for concreteness, $15=d^4-1$). Exact evolution of purity averaged over 2-site random gates is therefore determined by Markovian dynamics on this $d^{2n}$ dimensional operator space, with a 2-site matrix having a block diagonal structure, one block of size $1$ having the single matrix element equal to $1$, and the other $15\times 15$ block having all matrix elements equal to $1/15$, see Refs.~\cite{Oliveira,PRA08} for details. The 2-site matrix is therefore doubly stochastic and the process is evidently Markovian.

The dimensionality if this formulation can though be reduced down to dimension $2^n$~\cite{PRA08} irrespective of $d$, with the resulting 2-site matrix $M_{j,k}$ being equal to the Hamiltonian of a XY spin chain, $M_{j,k}=\frac{1}{2}(\mathbbm{1}\otimes\mathbbm{1}+h_{\rm XY})$, with $h_{\rm XY}=\frac{1+\gamma}{2} \sx{j}\sx{k}+\frac{1-\gamma}{2} \sy{j}\sy{k}+\frac{h}{2}(\sz{j}+\sz{k})$, with $\gamma$ and $h$ depending on $d$. For instance, for qubits ($d=2$; in the following we shall for concreteness always use $d=2$, the form of all matrices though stays the same for any $d$) one has~\cite{PRA08}
\begin{eqnarray}
\label{eq:Mpra}
	M_{j,k}&=&\begin{pmatrix}
		9/10 & 0 & 0 & 3/10 \\
		0 & 1/2 & 1/2 & 0 \\
		0 & 1/2 & 1/2 & 0 \\
		3/10 & 0& 0& 1/10
	\end{pmatrix}=\\
        &=&\frac{1}{2}\left(\mathbbm{1}\otimes \mathbbm{1}+\frac{4}{5}\sx{j} \sx{k}+\frac{1}{5}\sy{j} \sy{k}+\frac{2}{5}(\sz{j}+\sz{k})\right). \nonumber
\end{eqnarray} 
It is symmetric and conserves the parity of the number of 1s in the bit string.

Choosing a different basis one can also get a non-symmetric formulation with
\begin{equation}
	M_{j,k}=\begin{pmatrix}
		1 & 0 & 0 & 0 \\
		0 & 1/5 & 1/5 & 1/5 \\
		0 & 1/5 & 1/5 & 1/5 \\
		0 & 3/5& 3/5& 3/5
	\end{pmatrix}.
        \label{eq:Mnon}
\end{equation}
This matrix has only two nonzero eigenvalues, both equal to $1$, and is therefore a sum of two projectors (non-orthogonal) giving spectral decomposition,
\begin{equation}
  M_{j,k}=\ket{(1,0,0,0)}\bra{(1,0,0,0)}+\ket{\frac{(0,1,1,3)}{5}}\bra{(0,1,1,1)},
\end{equation}
where we used a bra-ket notation of a direct product of vectors. One can also write a SVD decomposition. Because in the above spectral decomposition the left eigenvectors happen to be orthogonal (which is in general not the case for non-symmetric matrices) the SVD decomposition is in fact, up-to different normalization of vectors, the same as the spectral decomposition, therefore
\begin{eqnarray}
  M_{j,k}=&1&\ket{(1,0,0,0)}\bra{(1,0,0,0)}+\\
  +&\frac{\sqrt{33}}{5}&\ket{\frac{1}{\sqrt{11}}(0,1,1,3)}\bra{\frac{1}{\sqrt{3}}(0,1,1,1)} \nonumber,
\end{eqnarray}
with the two singular values being $\sigma_1=1$ and $\sigma_2=\sqrt{33}/5$. The products of such 2-site matrices in the order of applied gates is then the matrix $M$ needed for evaluating the average purity at time $t$. For the above matrix (\ref{eq:Mnon}) the initial state is $\mathbf{x}(0)=(1,1,\ldots,1)$, i.e., $\ket{x(0)}=(\ket{0}+\ket{1})^{\otimes n}$. Calculating $\ket{x(t)}=M^t \ket{x(0)}$, purity is for a half-half bipartition given by
\begin{equation}
  I(t)=\frac{1}{N_A}\sum_{j=0}^{N_A-1} x_j(t)=\frac{1}{N_A} \braket{y}{x(t)},
  \label{eq:Isum}
\end{equation}
where $\ket{y}=(\ket{0}+\ket{1})^{\otimes n/2}\ket{0...0}$ and $N_A=2^{n/2}$. For other bipartitions one uses an obvious generalization: in $\ket{y}$ one takes $\ket{0}$ for all sites in subsystem $B$ and $(\ket{0}+\ket{1})$ for those in A. Summation in (\ref{eq:Isum}) comes from having to sum over squares of all state expansion coefficients that have $\mathbbm{1}$ on sites in the subsystem B. The steady state approached starting from $x(0)$ is $x(\infty)=(1,0,\ldots)+\frac{2^n-1}{4^n-1}(\ldots,3^{w(\mathbf{s})},\ldots)$, where $w(\mathbf{s})$ is the number of 1s in the bit string.

\subsection{Propagating purities}

In Ref.~\cite{Kuo20} a different procedure is used. Instead of propagating expansion coefficients one directly propagates an abstract vector encoding purity for all bipartitions. The resulting 2-site matrix is in this case
\begin{equation}
	M_{j,k}=\begin{pmatrix}
		1 & 0 & 0 & 0 \\
		\alpha & 0 & 0 & \alpha \\
		\alpha & 0 & 0 & \alpha \\
		0 & 0& 0& 1
	\end{pmatrix},
        \label{eq:Mkuo}
\end{equation}
with $\alpha=d/(d^2+1)$. Its spectral decomposition is a sum of two projectors,
\begin{equation}
  M_{j,k}=\ket{r_1}\bra{l_1}+\ket{r_2}\bra{l_2},
\end{equation}
with right and left eigenvectors $r_1=(1,0,0,-1),l_1=(1,0,0,0)$ and $r_2=(0,\alpha,\alpha,1),l_2=(1,0,0,1)$ (written in the 2-site basis $s_js_k$ ordered as $\{00,10,01,11 \}$, and normalization $\braket{r_j}{l_j}=1$, $\braket{l_1}{r_2}=\braket{l_2}{r_1}=0$).

The initial state is $\mathbf{x}(0)=(1,1,\ldots,1)$, i.e., $\ket{x}=(\ket{0}+\ket{1})^{\otimes n}$, $\mathbf{x}(t)=M^t \mathbf{x}(0)$, and half-half bipartite purity
\begin{equation}
I_{n/2}(t)=[\mathbf{x}]_{2^{n/2}-1}(t)=\braket{1\cdots 10 \cdots 0}{x(t)},
\end{equation}
with indices of $\mathbf{x}$ going from $0,1,..,2^n-1$.

The 2-site $M_{j,k}$ as well as the whole $M$ have a spin-flip symmetry, $X=\prod_j \sx{j}$, meaning that if $M \mathbf{y}=\lambda \mathbf{y}$ then also $X \mathbf{y}$ is an eigenvector with the same $\lambda$. Eigenvectors have a good parity $X$, they are either even or odd. For the relevant initial state $\mathbf{x}(0)$ only the even parity sector matters.

The above 2-site matrix (\ref{eq:Mkuo}) is not symmetric, however, by a similarity transformation one can transform it to a symmetric version. Namely, taking a single-site transformation $A_1$
\begin{equation}
A_1=\begin{pmatrix}
\sqrt{3} & 1\\
\sqrt{3} & -1
\end{pmatrix},
\end{equation}
and making a 2-site transformation $A=A_1 \otimes A_1$, one can transform the non-symmetric $M_{j,k}$ (\ref{eq:Mkuo}) to the symmetric $\tilde{M}_{j,k}=A^{-1}M_{j,k}A$, with
\begin{equation}
	\tilde{M}_{j,k}=\begin{pmatrix}
		9/10 & 0 & 0 & 3/10 \\
		0 & 1/2 & 1/2 & 0 \\
		0 & 1/2 & 1/2 & 0 \\
		3/10 & 0& 0& 1/10
	\end{pmatrix}.
\end{equation}
We can see that it is the same as the matrix (\ref{eq:Mpra}) obtained in Ref.~\cite{PRA08} starting from a state rather than purity evolution.

There are therefore several equivalent formulations with which one can propagate average purity in time. We are not going to study the full spectral properties of $M$ in any detail because we will first derive a simpler matrix description. Let us just state that on the relevant even subspace the $2^{n-1}$-dimensional $M_{\rm even}$ has $2^{n/2-1}$ nonzero eigenvalues and the kernel (zero eigenvalue) of algebraic multiplicity $2^{n-1}-2^{n/2-1}$ and geometric $2^{n-2}$. The kernel of $M_{\rm even}$ is a direct sum of Jordan normal blocks of different dimensions: $2^{n/2-1}$ blocks of dimension $\frac{n}{2}$, $2^{n/2-1}$ blocks of dimension $\frac{n}{2}-1$, $2^{n/2}$ blocks of dimension $\frac{n}{2}-2$, $2^{n/2+1}$ blocks of dimension $\frac{n}{2}-3$, and so on (decreasing size by $1$ and increasing degeneracy by $2$), and finally $2^{n-3}$ blocks of dimension $1$.

\section{Calculating purity}
\label{App:B}

Let us use the 2-site matrix $M_{j,k}$ in Eq.(\ref{eq:Mkuo}), the staircase configuration,
\begin{equation}
 M=M_{n,n-1}\cdots M_{1,2},
\end{equation}
and for a starter calculate $I_{n/2}(t=1)$ for any system size $n$. This simple calculation will guide us towards more complicated results for larger $t$ and the recursion formula for $I_k(t)$.

To avoid cluttered indices let us denote the vector of all $2^n$ purities $I_\mathbf{s}$ by $x$ and label its components by bitstrings. The half-half bipartition purity is then $I_{n/2}(1)=x_{1\cdots 1|0\cdots 0}(1)$, where we use a vertical line to denote the bipartition cut. Vector $x$ is propagated as $x(t+1)=M x(t)$, and one starts with $x_\mathbf{s}(t=0)=1$.

Looking at $M_{j,k}$ (\ref{eq:Mkuo}) we see that we can get to state $00$ only from $00$, and therefore to be in the state $1\cdots1|0\cdots0$ at $t=1$ we already need to be in that very same state after the gate $M_{n/2-1,n/2}$ that acts across the cut is applied (all latter gates are benign because they can not change the $00$ on any pair of bits in B). Denoting by $y$ the state before the cut-gate is applied, we have $I_{n/2}(1)=x_{1\cdots 1|0\cdots 0}(1)=\alpha y_{1\cdots 10|0\cdots 0}+\alpha y_{1\cdots 11|10\cdots 0}$. Working backwards on gates applied on A, we can get to $11$ only from $11$, while one can get to $10$ from either $00$ or $11$. Each $10$ term can be obtained from 2 possible bit strings on the previous step. For $n=6$ we for instance have $I_3(1)=\alpha x_{111|100}(0)+\alpha (\alpha x_{111|000}(0)+\alpha^2 (x_{000|000}(0)+x_{110|000}(0)))$. For general $n$ one has
\begin{eqnarray}
I_{n/2}(1)&=&\alpha x_{\bar{1}1|1\bar{0}}+\alpha^2 x_{\bar{1}1|0\bar{0}}+\alpha^3 x_{\bar{1}0|0\bar{0}}+\cdots +\nonumber \\
&&+\alpha^{n/2} x_{11\bar{0}|\bar{0}}+\alpha^{n/2} x_{00\bar{0}|\bar{0}}=\nonumber \\
&=& \alpha^{n/2} x_{\bar{0}|\bar{0}}+\sum_{r=2}^{n/2+1} \alpha^{n/2+2-r} x_{1_1\cdots 1_r \bar{0}},
\label{eq:Ix}
\end{eqnarray}
where we use $\bar{0}$ and $\bar{1}$ for a number of consecutive repeated bits (their number is such that the number of all bits in a subsystem is $n/2$), and $x_{1_1\cdots 1_r \bar{0}}$ is a domain-wall with $r$ 1s. Using the initial vector gets us
\begin{equation}
I_{n/2}(1)=\alpha^{n/2}+\sum_{k=1}^{n/2} \alpha^k=\frac{\alpha}{1-\alpha}+\frac{1-2\alpha}{1-\alpha} \alpha^{n/2}.
\label{eq:Isum0}
\end{equation}
Specifically, for $d=2$
\begin{equation}
I_{n/2}(1)=\frac{2}{3}+\frac{1}{3}\left(\frac{2}{5}\right)^{n/2},
\end{equation}
and for $d=3$
\begin{equation}
I_{n/2}(1)=\frac{3}{7}+\frac{4}{7}\left(\frac{3}{10}\right)^{n/2}.
\end{equation}

That was for a half-cut. By similar arguments one can also write $I_k(1)$ for a cut after 1st $k$ consecutive spins in A -- one just has to replace $n/2$ in eq.(\ref{eq:Isum0}) by $k$ (all three occurrences). Purity for a cut after $k$ spins is therefore,
\begin{equation}
I_k(1)=\frac{\alpha}{1-\alpha}+\frac{1-2\alpha}{1-\alpha} \alpha^{k}.
\label{eq:Ik1}
\end{equation}
This is an exact expression holding for any $n$.

We also immediately recognize that the coefficients $x_j$ in Eq.(\ref{eq:Ix}) are nothing but $I_k$, so that we can write a recursion
\begin{equation}
  I_k(t)=\alpha^k+\sum_{r=2}^{k+1} \alpha^{k+2-r} I_r(t-1).
  \label{eq:Irec}
\end{equation}
This recursion is a significant simplification of the original $x(t+1)=Mx(t)$ iteration as one does not have to deal with an exponentially large $M$, but rather just $\sim n$ relevant continuous-A purities.

The recursion (\ref{eq:Irec}) can be used to obtain purity at later times. For instance, for a half-cut we obtain
\begin{equation}
I_{n/2}(2)=\left(\frac{\alpha}{1-\alpha}\right)^2+\alpha^{n/2} \frac{1-2\alpha}{1-\alpha} \left(\frac{1}{1-\alpha}+\alpha^2 \frac{n}{2}\right),
\end{equation}
where we need to have $n\ge 4$ in order for all $I_k(1)$ used in the recursion to make sense (be defined and nonzero). For $d=2$ the term in the last bracket is $\frac{5}{3}+\frac{2n}{25}$.
For general $k$ we get
\begin{equation}
I_{k}(2)=\left(\frac{\alpha}{1-\alpha}\right)^2+\alpha^{k} \frac{1-2\alpha}{1-\alpha} \left(\frac{1}{1-\alpha}+\alpha^2 k\right),
\end{equation}
Iterating again, we get the result at $t=3$,
\begin{eqnarray}
&&I_{k}(3)=\left(\frac{\alpha}{1-\alpha}\right)^3+\\
&&+\alpha^{k} \frac{(1-2\alpha)[(1-\alpha\beta)+k\alpha^2\beta(1+3\alpha^2\beta/2)+k^2 \alpha^4\beta^2/2]}{\beta^3}, \nonumber
\end{eqnarray}
where $\beta \equiv 1-\alpha$. For instance, for $k=n/2$ and $d=2$ we have
\begin{equation}
I_{n/2}(3)=\left(\frac{2}{3}\right)^3+\left(\frac{2}{5}\right)^{n/2}\frac{1}{3^3 5^2}\left(475+\frac{858}{25}n+\frac{18}{25}n^2\right),
\end{equation}
holding for $n\ge 6$.

It is clear from these explicit results as well as from the recursive formula that the general result is of the form
\begin{equation}
I_{n/2}(t)=\left(\frac{\alpha}{1-\alpha}\right)^t+\alpha^{n/2} p_{t-1}(n),
\end{equation}
holding for $n \ge 2t$, and where $p_k(x)$ is a polynomial of order $k$ in $x$. This means that in the TDL one has at any finite $t$ purity $I_k(t)=(\alpha/(1-\alpha))^t$ for any extensive $k \propto n$. The phantom eigenvalue giving this thermodynamically relevant decay is therefore
\begin{equation}
  \lambda_{\rm ph}=\frac{\alpha}{1-\alpha}=\frac{d}{d(d-1)+1},
  \label{eq:lph}
\end{equation}
and is $\lambda_{\rm ph}=\frac{2}{3}$ for $d=2$, $\lambda_{\rm ph}=\frac{3}{7}$ for $d=3$, while for $d \gg 1$ it scales as $\lambda_{\rm ph} \asymp \frac{1}{d}$.

\section{Matrix formulation}
\label{App:C}

We could also write purity recursion (\ref{eq:Irec}) as a matrix iteration. Defining an $n$-component vector of purities $\mathbf{\tilde I}=(1,I_2,I_3,\ldots,I_{n-1},1)$, one has $\mathbf{\tilde I}(t+1)=A \mathbf{\tilde I}(t)$, where the $n\times n$ matrix $A$ is
\begin{equation}
A=\begin{pmatrix}
    1 & 0 & 0 \\
   \mathbf{a}_1 & T & \mathbf{a}_2 \\
   0 & 0 & 1
\end{pmatrix},
\label{eq:A}
\end{equation}
with the vector $\mathbf{a}_1=(\alpha^2,\alpha^3,\ldots,\alpha^{n-1})$, $\mathbf{a}_2=(0,\ldots,0,\alpha)$, and $(n-2)\times (n-2)$ Toeplitz matrix $T$
\begin{equation}
T=\begin{pmatrix}
    \alpha^2 & \alpha & 0 & \cdots & 0 \\
    \alpha^3 & \alpha^2 & \alpha & \cdots & 0\\
    \vdots   &  \vdots & \ddots & \cdots & \vdots \\
    \alpha^{n-2} & \alpha^{n-3} & \ddots & \ddots & \alpha\\
    \alpha^{n-1} & \alpha^{n-2} & \cdots & \cdots & \alpha^2 
\end{pmatrix}.
\label{eq:Ts}
\end{equation}
Alternatively, defining $\mathbf{I}=(I_2,I_3,\ldots,I_{n-1})$ we could write $\mathbf{I}(t+1)=\mathbf{a}+T\mathbf{I}(t)$, where $\mathbf{a}=\mathbf{a}_1+\mathbf{a}_2$. This matrix scheme gives the exact purity for any $k$, $n$ and $t$ (in the above formulas for $I_k(t=1,2,3)$ holding for $n\ge 2t$ we evaluated the sums as if $A$ would be of infinite size, i.e., no boundary effects; this matrix formulation though correctly accounts also for boundaries).

We remark that the transpose $M^{\rm T}$ preserves a subspace of domain wall bit strings, that is the subspace spanned by $\{\bar{0},11\bar{0},111\bar{0},\ldots,\bar{1}\}$ is invariant under $M^{\rm T}$. Similarly, also this subspace with bits flipped is invariant. So one can form an invariant basis of $n-1$ 1-domain-wall states even under spin flip, $\{\bar{0}+\bar{1},11\bar{0}+00\bar{1},\ldots \}/\sqrt{2}$. Projection of $M^{\rm T}$ to this subspace would give a $(n-1)$ dimensional matrix that is essentially the same as $A$ (\ref{eq:A}). One could also extend this projection to $r$-domain-wall states. This was used in Ref.~\cite{Kuo20} to discuss decay of purity. From such projections of $M^{\rm T}$ one could obtain the corresponding left eigenvectors of $M$.

\begin{figure}[t!]
\centerline{\includegraphics[width=.8\columnwidth]{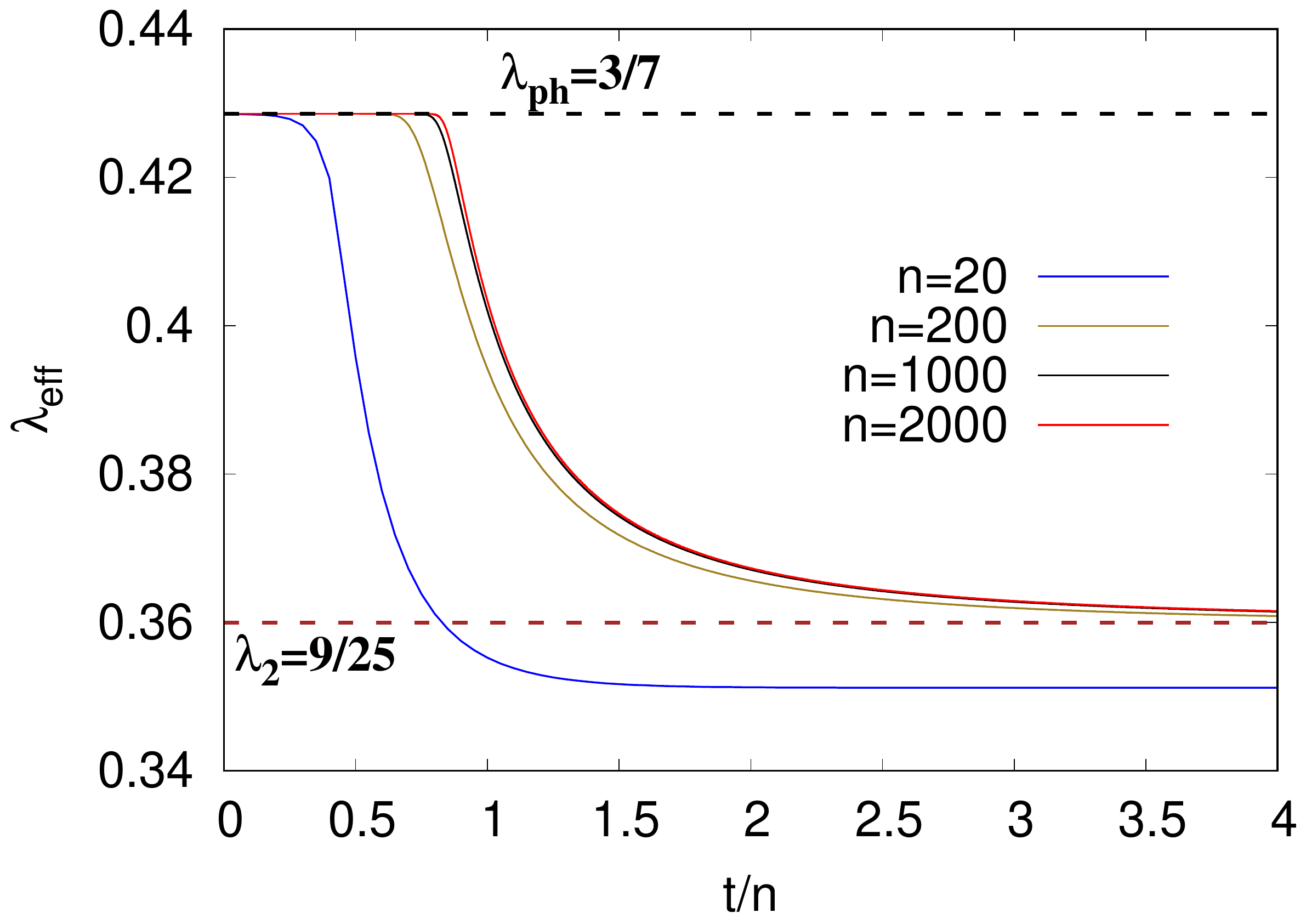}}
\caption{(Color online) Effective purity decay rate $\lambda_{\rm eff}=\exp{(I'/I)}$, i.e., $I(t)=C \lambda_{\rm eff}^t$, for qutrits $d=3$.}
\label{fig:lambdad3}
\end{figure}

Because $A$ is of size $n$ we can easily calculate exact average purity dynamics for thousands of sites. In Fig.~\ref{fig:lambdad3} we show how the local effective rate changes with time, thereby more clearly seeing the transition from the initial decay $I(t)=\lambda_{\rm ph}^t$ with the phantom $\lambda_{\rm ph}$ to the asymptotic decay $I(t) \asymp \lambda_2^t$ with the 2nd largest eigenvalue of $M$. We define $I(t)=C \lambda_{\rm eff}^t$, from which one can calculate $\lambda_{\rm eff}=\exp{[I'(t)/I(t)]}$ where $I'(t)=dI/dt$.
We can see (Fig.\ref{fig:lambdad3}) that the transition time between the two rates is proportional to $n$, and therefore diverges in the TDL limit.

\begin{figure*}[t]
\centerline{\includegraphics[width=.64\columnwidth]{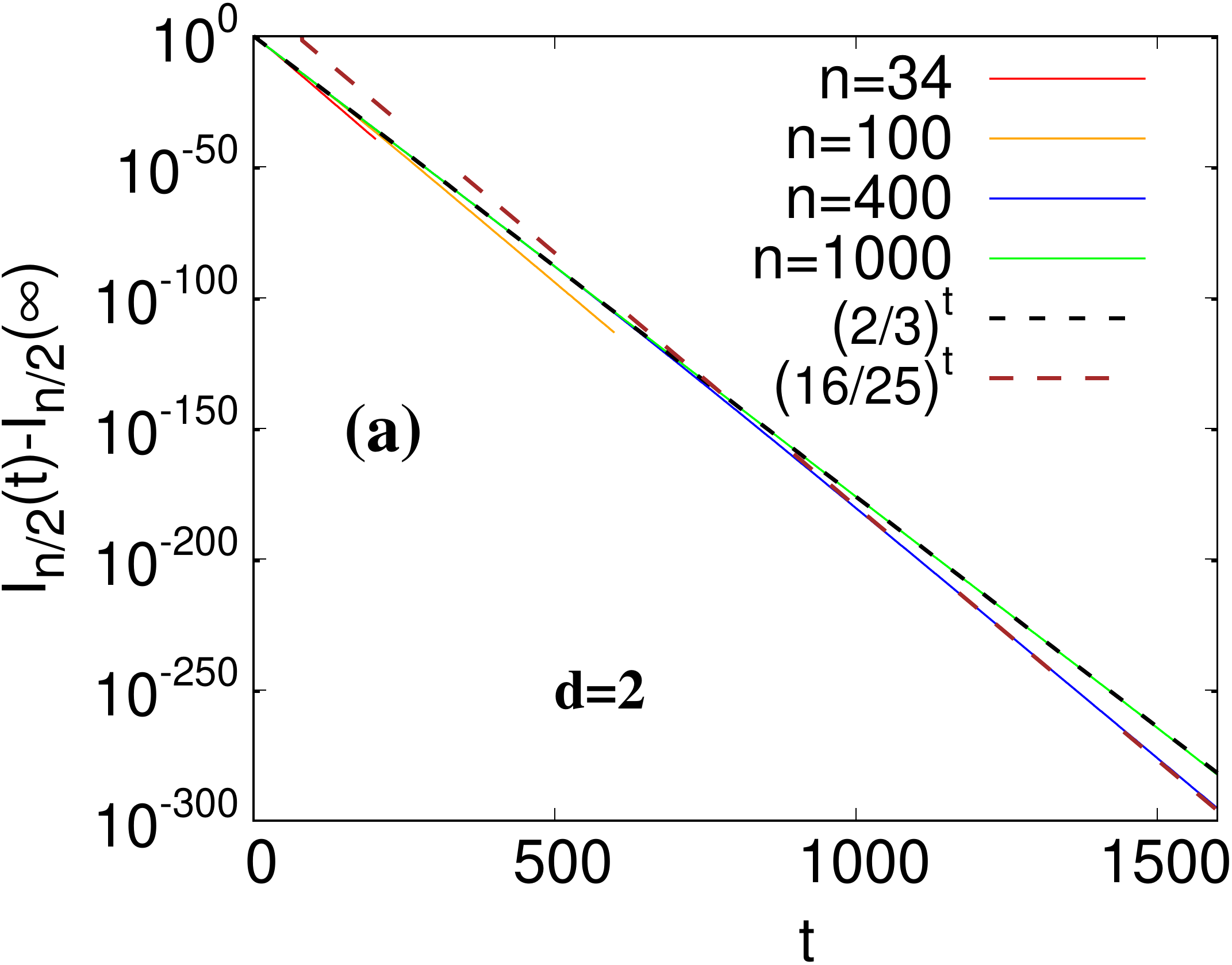}\hskip3pt\includegraphics[width=.62\columnwidth]{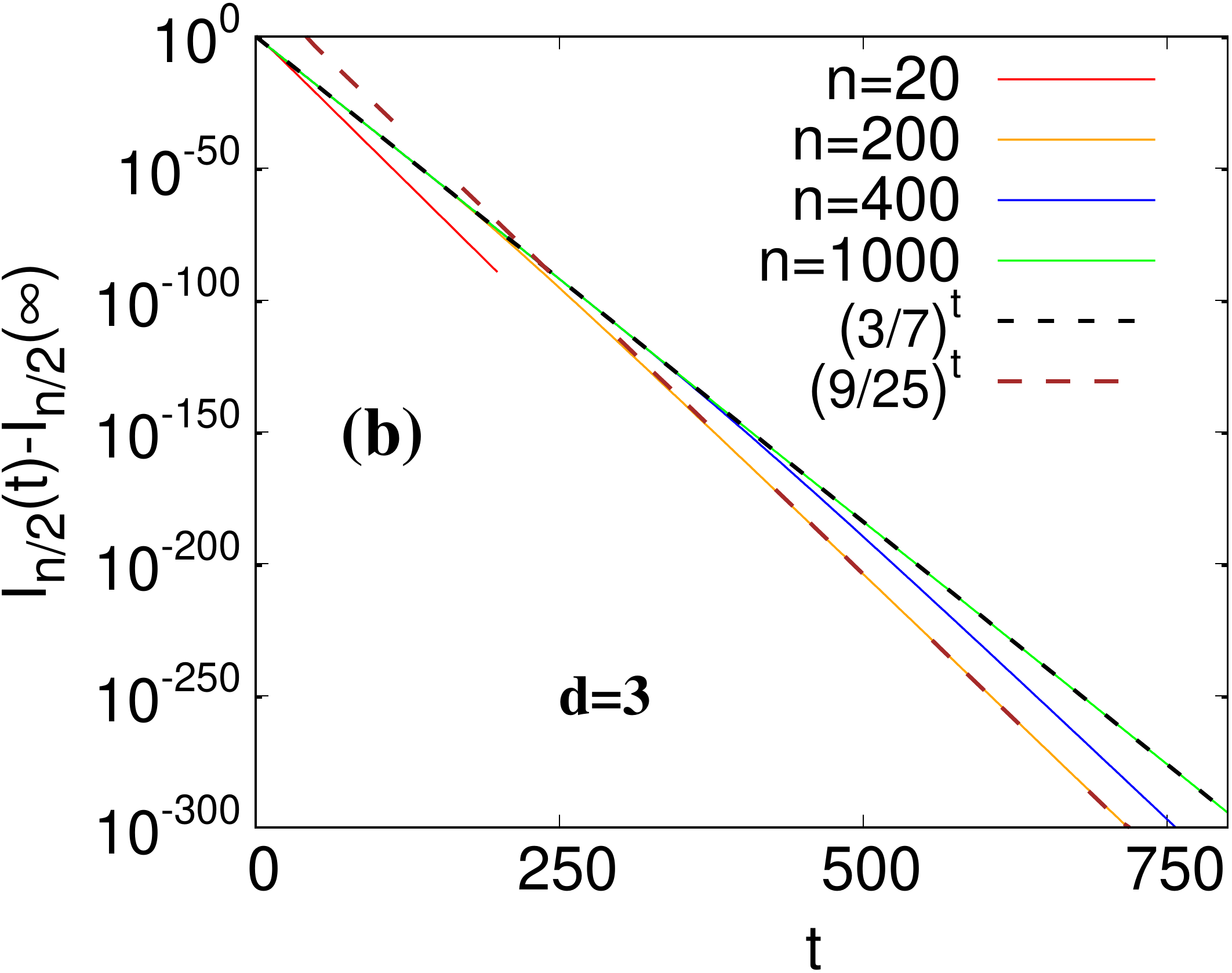}\hskip3pt\includegraphics[width=.64\columnwidth]{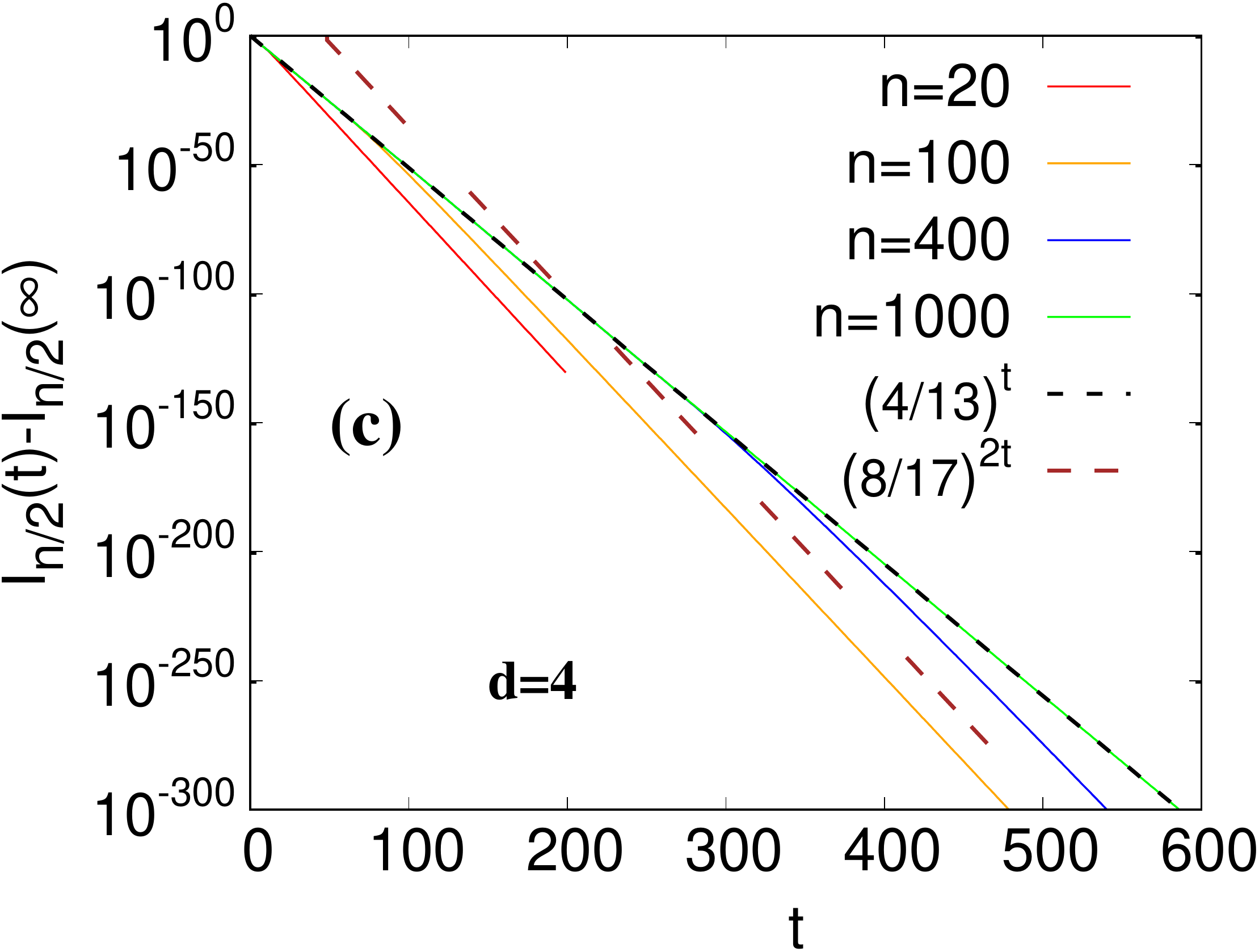}}
\caption{(Color online) Purity decay for qubits (a), qutrits (b), and $d=4$ (c). One can see agreement with $\lambda_{\rm ph}$ (\ref{eq:lph}) for times smaller than $\sim n$, and asymptotic decay with $\lambda_2$ (\ref{eq:l2}) for longer times.}
\label{fig:I234}
\end{figure*}

\subsection{Spectral decomposition}
Crucial is matrix $T$ which has a form of an asymmetric hopping on a lattice of $n-2$ sites, with the left hops being of length $1$ only, while all right hops are allowed. Toeplitz matrix $T$ has a Hessenberg form. By a similarity transformation it can be transformed to a matrix where all the nonzero matrix elements are, instead of being powers of $\alpha$, equal
\begin{equation}
R^{-1} T R=\alpha^2 \begin{pmatrix}
    1 & 1 & 0 & \cdots & 0 \\
    1 & 1 & 1 & \cdots & 0\\
    \vdots   &  \vdots & \ddots & \cdots & \vdots \\
    1 & 1 & \ddots & \ddots & 1\\
    1 & 1 & \cdots & \cdots & 1 
\end{pmatrix},
\end{equation}
where one uses diagonal $R={\rm diag}(\alpha,\alpha^2,\ldots,\alpha^{n-2})$.

Eigenvalues of $T$ (\ref{eq:T}) are roots of the characteristic polynomial $\det{(T-\tilde{\lambda} \mathbbm{1})}=0$. One can get a 2-step recursive relation in $n$ for the determinant by subtracting from the first column the 2nd column multiplied by $\alpha$. The final expression for the characteristic equation is~\cite{Fairweather71} $D_n=\det{(T-\tilde{\lambda} \mathbbm{1})}=(-1)^n\alpha^{n-1} \tilde{\lambda}^{(n-3)/2}U_{n-1}(\sqrt{\tilde{\lambda}/4}/\alpha)=0$, where $U_k$ is the Chebyshev polynomial of the second kind, $U_1(x)=2x, U_2(x)=4x^2-1,\ldots$, or explicitly $U_k(\cos{\varphi})=\sin{[(k+1)\varphi]}/\sin{\varphi}$.

One therefore has eigenvalue $\tilde{\lambda}=0$ with algebraic multiplicity $n/2-1$ and geometric multiplicity $1$, and $n/2-1$ nonzero eigenvalues (assuming $n$ even)
\begin{equation}
  \tilde{\lambda}_j =4\alpha^2 \cos^2{\left(\frac{j\pi}{n}\right)}  ,\quad j=1,\ldots,n/2-1.
  \label{Aeq:lj}
\end{equation}
The kernel therefore has just one Jordan block of size $n/2-1$ while all other eigenvalues are real and non-degenerate for any finite $n$.

As a side remark: if one would take in $T$ only the superdiagonal such a matrix would have $(n-2)$ dimensional Jordan block with $\lambda=0$; taking the diagonal and the superdiagonal one would again have only one $(n-2)$ dimensional Jordan block with $\lambda=\alpha^2$; while deleting the superdiagonal (triangular matrix) one would have a $(n-2)$ dimensional Jordan block with $\lambda=\alpha^2$. $T$ is therefore in a way a simple solvable deformation of a single Jordan block resulting in a non-trivial spectrum.

Denoting the corresponding right eigenvector as $\ket{R_j}$ and writing $\varphi_j:=j \pi/n$, its components are
\begin{equation}
  [R_j]_k= (2\alpha \cos{\varphi_j})^{k-2} U_k(\cos{\varphi_j}),\quad k=1,\ldots,n-2.
\end{equation}
The left eigenvector $\ket{L_j}$ corresponding to the same eigenvalue $\tilde{\lambda}_j$ can be obtained simply by reflecting vector components, i.e., replacing index $k \to n-1-k$, or explicitly
\begin{equation}
  [L_j]_k= (2\alpha \cos{\varphi_j})^{n-3-k} U_{n-1-k}(\cos{\varphi_j}),\quad k=1,\ldots,n-2.
\end{equation}

\subsection{Decomposition of $A$}

Once we have the spectral decomposition of $T$ it is easy to write down the spectral decomposition of $A$. The spectrum of $A$ is a union of the spectrum of $T$ plus two eigenvalues $\lambda_1=1$, one in the even and one in the odd sector. All eigenvalues of $A$ are also eigenvalues of $2^n$ dimensional $M$ on the space of all bipartite purities. In fact, the largest nontrivial (i.e., smaller than $1$) eigenvalue of $M$ is $\tilde{\lambda}_1$, and therefore we have an exact expression for the 2nd largest eigenvalue $\lambda_2$ of $M$,
\begin{equation}
  \lambda_2=4\alpha^2 \cos^2{\left( \frac{\pi}{n}\right)} \longrightarrow \frac{4d^2}{(d^2+1)^2},
  \label{eq:l2}
\end{equation}
that determines the eventual asymptotic decay of purity, that is, at any fixed $n$ and for large enough $t$ one has $I_\mathbf{s}(t)-I_\mathbf{s}(\infty) \sim \lambda_2^t$. For instance, one has $\lambda_2=\frac{16}{25},\frac{9}{25},\left(\frac{8}{17}\right)^2$, for $d=2,3,4$, respectively. In Fig.~\ref{fig:I234} we show purity relaxation for $d=2,3,4$. We can see that the transition point between the initial phantom decay $\lambda_{\rm ph}^t$ and the asymptotic $\sim \lambda_2^t$ moves to infinity with increasing system size $n$. Also worth noting is that with increasing $d$ the ratio $\lambda_{\rm ph}/\lambda_2 \asymp \frac{d}{4}$ increases, and therefore the transition becomes more prominent.

Going to eigenvectors of $A$, the even left and right eigenvector corresponding to the eigenvalue $1$ form a projector to the steady state, $\ket{R}\bra{L}$, and have components $R_k=I_k(\infty)=(d^k+d^{n-k})/(1+d^n)$ for $k=2,\ldots,n-1$ and $R_1=R_n=1$, while $L=(\frac{1}{2},0,\ldots,0,\frac{1}{2})$. Eigenvectors of $A$ corresponding to nonzero eigenvalues $\tilde{\lambda}_j$ (\ref{eq:lj}) can be constructed from those of $T$. Denoting them by $\tilde{R}_j$, one has $[\tilde{R}_j]_{k=1,n}=0$ and $[\tilde{R}_j]_{k}=[R_j]_{k-1}$ for $k=2,\ldots,n-1$. For the left eigenvectors one has instead $[\tilde{L}_j]_{k}=[L_j]_{k-1}$ for $k=2,\ldots,n-1$, while $[\tilde{L}_j]_{1}=\braket{L_j}{\mathbf{a}_1}/(\tilde{\lambda}_j-1)$ and $[\tilde{L}_j]_{n}=\braket{L_j}{\mathbf{a}_2}/(\tilde{\lambda}_j-1)$.

Once we normalize left and right eigenvectors as $\braket{\tilde{L}_j}{\tilde{R}_j}=1$, it would be tempting to write the relevant spectral decomposition for $A$ as $A=\ket{R}\bra{L}+\sum_{j=1}^{n/2-1} \tilde{\lambda}_j \ket{\tilde{R}_j}\bra{\tilde{L}_j}$, however, that would give correct purity only at later times. For instance, for $I_{n/2}(t)$ we would get the correct value only for $t \ge n/4$ (while $I_2(t)$ would be correct only for $t \ge n/2-1$). This, at first sight surprising failure, is due to the Jordan normal form of the kernel. Remembering that the single Jordan normal block is of size $n/2-1$, we have the spectral decomposition $A=P d P^{-1}$, where $d$ has the Jordan normal form
\begin{equation}
d=\left( \begin{array}{ccccc|ccc}
    0 & 1 &   &   &  & & & \\
      & 0 & 1 &   &  & & & \\
      &   & \ddots & \ddots & & && \\
    &   &        & 0 & 1 &  & &\\
    &   &        &  & 0 &  & &\\
    \hline
    &   &        & &  & 1 & &\\
    &   &        & &  & & \tilde{\lambda}_1 & \\
    &   &        &  &  & & &\ddots 
\end{array}
\right).
\end{equation}
Rows of $P^{-1}$ are left eigenvectors, columns of $P$ are right eigenvectors of $A$. Therefore one has
\begin{equation}
A=\ket{R}\bra{L}+\sum_{j=1}^{n/2-1} \tilde{\lambda}_j \ket{\tilde{R}_j}\bra{\tilde{L}_j}+\sum_{k=1}^{n/2-2} \ket{r_k}\bra{l_{k+1}},
\end{equation}
where we have separated the kernel part with eigenvectors $l_k$ and $r_k$ that satisfy $A^{p\ge k} \ket{r_k}=0$, and $(A^T)^{n/2-k} \ket{l_k}=0$ and $A^T \ket{l_p}=\ket{l_{p+1}}$. In other words, $\ket{l_1}$ is in the kernel of $(A^T)^{n/2-1}$ but not in the kernel of any lower power of $A^{\rm T}$, see e.g. Ref.~\cite{Weintraub} for a summary of Jordan normal forms. Namely, because the geometrical multiplicity of $\lambda=0$ is 1 there are vectors $\mathbf{y}$ such that $A\mathbf{y} \neq 0$, but $A^r \mathbf{y}=0$ for some $1<r\le n/2-1$.

\begin{figure}[t!]
\centerline{\includegraphics[width=.9\columnwidth]{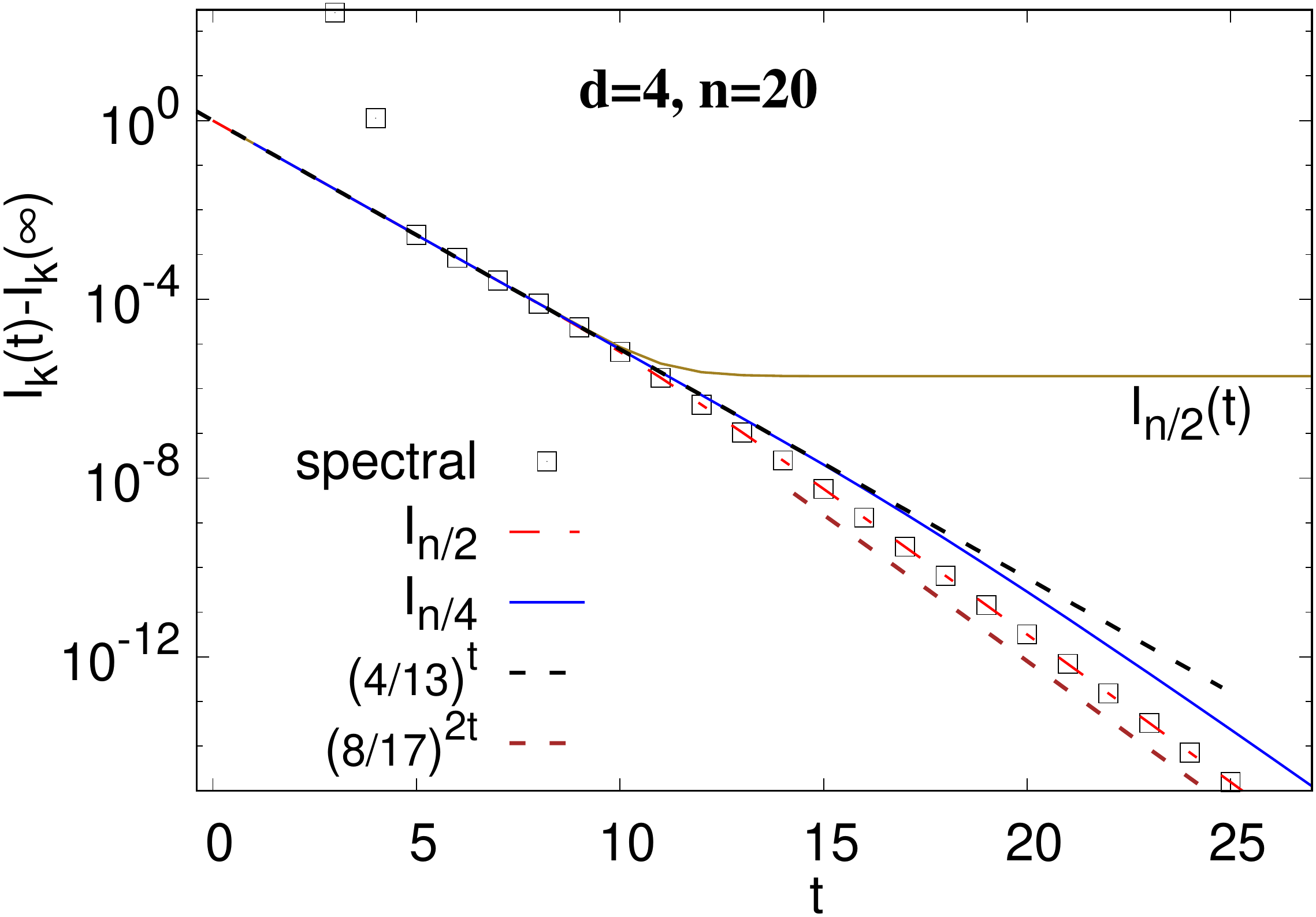}}
\caption{(Color online) Purity decay for $d=4$ for a half-cut, $I_{n/2}$, and for a quarter cut $I_{n/4}$. The saturating curve shows $I_{n/2}(t)$ without subtracting $I_{n/2}(\infty)$, while black squares are using spectral decomposition with only nonzero eigenvalues of $A$, thereby not correctly accounting for its Jordan kernel structure.}
\label{fig:u16}
\end{figure}

Making powers of $A$ the size of the kernel shrinks, for instance for $A^2$ we have
\begin{equation}
A^2=\ket{R}\bra{L}+\sum_{j=1}^{n/2-1} \tilde{\lambda}_j^2 \ket{\tilde{R}_j}\bra{\tilde{L}_j}+\sum_{k=1}^{n/2-3} \ket{r_k}\bra{l_{k+2}},
\end{equation}
i.e., the kernel part shifts by $2$ instead of by $1$ as in $A$. In $A^{n/2-1}$ one has no kernel part anymore and all evolution is correct just using nonzero $\tilde{\lambda}_j$ in the spectral decomposition. This is the reason that using only nonzero eigenvalues does not correctly capture the initial decay. We illustrate this in Fig.~\ref{fig:u16} where one can see that the spectral decomposition with only $\tilde{\lambda}_j$ terms gives correct $I_{n/2}$ only for $t \ge n/4=5$. We also show purity for a bipartition into 1st $n/4$ sites and the rest. One can see that, as predicted, $I_{n/4}$ also exhibits the transition in the decay rate. In fact, the initial phantom relaxation $\lambda_{\rm ph}^t$ holds even upto slightly longer times than for $I_{n/2}$.

\subsection{Brick-wall configuration}

An interesting observation is that for a brick-wall configuration of gates with open boundary conditions the decay of purity is given by the same $\lambda_2$ (\ref{eq:l2}) as calculated here, however, without the phantom relaxation~\cite{PRX21}. Such asymptotic decay $\lambda_2^t$ for the random brick-wall protocol has been derived before~\cite{Adam18,Frank18,Kuo20}. The fact that $\lambda_2$ is the same for staircase and brick-wall configurations is a consequence of the exact spectral equivalence of all configurations with open boundary conditions, see Ref.~\cite{PRX21}. It might be instructive though to see this using the same techniques as used here for the staircases. Looking again at $I_k$, one notices that purities for even-$k$ consecutive sites decouple from odd $k$. One can write recursive equations just for $I_k$ with even $k$. Taking again even $n$, and defining purity vector $\mathbf{I}=(I_2,I_4,\ldots,I_{n-2})$, having $n/2-1$ components, one gets iteration $\mathbf{I}(t+1)=\mathbf{a}+T\mathbf{I}(t)$, where $\mathbf{a}=(\alpha^2,0,\ldots,0,\alpha^2)$ and
\begin{equation}
T=\alpha^2\begin{pmatrix}
    2 & 1 &  &  \\
    1 & 2 & 1 &  \\
       & \ddots & \ddots & \ddots & \\
       & & 1 & 2 & 1 \\
       & & & 1 & 2 \\
\end{pmatrix}.
\label{eq:Tbw}
\end{equation}
$T$ is a tridiagonal Toeplitz matrix. Its exact eigenvalues are exactly the same $\tilde{\lambda}_j$ as for the staircases configuration (\ref{eq:lj}). Furthermore, $T$ is symmetric, i.e., normal, its symbol $a(z)$ is real on the unit circle, and the spectra of the operator (infinite matrix) and of a finite matrix in the limit $n \to \infty$ coincide, it fills the interval $[ 0,4\alpha^2]$. Therefore, there is no phantom relaxation for the brick-wall configuration with open boundaries.

\newpage
\clearpage

\setcounter{equation}{0}
\setcounter{figure}{0}
\setcounter{table}{0}
\makeatletter
\renewcommand{\theequation}{\thesection.\arabic{equation}}
\renewcommand{\thefigure}{\thesection.\arabic{figure}}


\end{document}